\documentclass[aps,prl,twocolumn,superscriptaddress,showpacs]{revtex4}

\usepackage{amsmath}
\usepackage{bm}
\usepackage{graphicx}

\begin{document}

\title{Anderson transition in low-dimensional disordered systems driven by
nonrandom long-range hopping}

\author{A.\ Rodr\'{\i}guez}
\affiliation{Departamento de Matem\'{a}tica Aplicada y Estad\'{\i}stica,
Universidad Polit\'{e}cnica, E-28040 Madrid, Spain} 

\author{V.\ A.\ Malyshev}
\affiliation{Departamento de F\'{\i}sica de Materiales, Universidad
Complutense, E-28040 Madrid, Spain} 
\thanks{On leave from ``S.I. Vavilov State Optical Institute'',
Saint-Petersburg, Russia.}

\author{G.\ Sierra}
\affiliation{Instituto de Matem\'aticas y F\'{\i}sica Fundamental, 
CSIC, Madrid, Spain} 

\author{M.\ A.\ Mart\'{\i}n-Delgado}
\affiliation{Departamento de F\'{\i}sica Te\'{o}rica I, 
Universidad Complutense,
E-28040 Madrid, Spain} 

\author{J.\ Rodr\'{\i}guez-Laguna}
\affiliation{Departamento de F\'{\i}sica Te\'{o}rica I, Universidad Complutense,
E-28040 Madrid, Spain} 

\author{F.\ Dom\'{\i}nguez-Adame}
\affiliation{Departamento de F\'{\i}sica de Materiales, Universidad
Complutense, E-28040 Madrid, Spain} 

\date{\today}

\begin{abstract}

The single-parameter scaling hypothesis predicts the absence of  delocalized
states for noninteracting quasiparticles in low-dimensional disordered systems.
We show analytically and numerically that extended states may occur in the one-
and two-dimensional Anderson model with a {\it nonrandom\/} hopping falling off
as some {\em power\/} of the distance between sites. The different size scaling
of the bare level spacing and the renormalized magnitude of the disorder seen
by the quasiparticles finally results in the delocalization of states at one of
the band edges of the quasiparticle energy spectrum. The delocalized nature of
those eigenstates is investigated by numerical diagonalization of the
Hamiltonian and by the supersymmetric method for disorder averaging, combined
with a renormalization group analysis.

\end{abstract}


\maketitle

Localization of noninteracting quasiparticles in random media with 
time-reversal symmetry and finite-range hopping have been extensively studied
since the seminal paper by Anderson~\cite{Anderson58}. The hypothesis of
single-parameter scaling, introduced in Ref.~\onlinecite{Abrahams79}, led to
the general belief that all eigenstates of noninteracting quasiparticles were
exponentially localized in one~(1D) and two~(2D) dimensions (see 
Refs.~\cite{Lee85,Kramer93} for a comprehensive review) and that
localization-delocalization transitions no longer exist in the  thermodynamic
limit. Despite models with finite-range hopping work nicely in describing a
variety of materials, long-range hopping are often found in different physical
systems (e.g.\@ Frenkel excitons). Random long-range hopping was found to give
rise to delocalization of states not only in three dimensional
systems~\cite{Anderson58} but in any
dimension~\cite{Logan87,Levitov89,Mirlin96,Parshin98}. Recent
studies~\cite{Deych00} revised the validity of the single-parameter scaling
hypothesis even within the original 1D Anderson model with nearest-neighbor
coupling, although did not question the statement that all eigenstates in 1D
random systems are localized.

In this Letter we present analytical and numerical proofs that a
localization-delocalization transition may occur in 1D and 2D systems with 
{\it diagonal\/} disorder and {\it nonrandom\/} intersite coupling which falls 
off according to a power-like law. Apart from the importance of this finding
from a general point of view, it may be relevant for several physical systems.
As an example, let us mention dipolar Frenkel excitons on 2D regular lattices
where molecules are subjected to randomness due to a disordered
environment~\cite{Nabetani95}. Biological light-harvesting antenna systems
represent a realization of the model we are dealing
with~\cite{Kopelman97,dendrimers}. Magnons in 1D and 2D disordered spin systems
provide one more example of interest.

We consider the Anderson Hamiltonian on a $d$-dimen\-sional ($d=1,2$) simple
lattice with ${\cal N} = N^d$ sites:
\begin{equation}
{\cal H} = \sum_{\bf n} \varepsilon_{\bf n} |{\bf n}\rangle \langle {\bf n}|
+\sum_{\bf {nm}} J_{\bf {nm}}|{\bf n}\rangle \langle {\bf m}| \ ,
\label{Hsite}
\end{equation}
where $|{\bf n} \rangle$ is the ket-vector of the state localized at site ${\bf
n}$, and $\{\varepsilon_{\bf n}\}$ are random site energies, assumed to be
uncorrelated for different sites and distributed uniformly within an interval
$[-\Delta/2,\Delta/2]$, thus having zero mean and standard deviation $\sigma =
\Delta/\sqrt{12}$. The hopping  integrals between lattice sites $\mathbf{m}$ and
$\mathbf{n}$ will be taken in the form $J_{\mathbf{mn}} = J/|\mathbf{m} -
\mathbf{n}|^{\mu} \quad (J_{\mathbf{mm}} \equiv 0)$, where $J > 0$ without loss
of generality  and the lattice constant is set to unity. We stress that hopping
integrals do not fluctuate.

For our qualitative reasoning we rewrite the Hamiltonian~(\ref{Hsite}) in the
Bloch wave representation, $|\mathbf{k} \rangle = {\cal N}^{-1/2} \sum_{\bf n}
\exp({i{\bf kn}})|{\bf n} \rangle$, with periodic boundary conditions. It then
reads 
\begin{subequations}
\label{1}
\begin{equation}
{\cal H} = \sum_{\bf k} E_{\bf k} |{\bf k}\rangle \langle {\bf k}|
+ \sum_{\bf {kk^\prime}} (\delta {\cal H})_{\bf {kk^\prime}}
|{\bf k}\rangle \langle {\bf k^\prime}| \ ,
\label{Hk}
\end{equation}
\begin{equation}
E_{\bf k} = J \sum_{{\bf n}\neq{\bf 0}} \>\frac{e^{i{\bf kn}}}{|\bf n|^\mu} \ ,
\label{Ek}
\end{equation}
\begin{equation}
(\delta {\cal H})_{\bf kk^\prime} = \frac{1}{\cal N}
\sum_{{\bf n}} \varepsilon_{\bf n} e^{i({\bf k - k^\prime) n}}\ ,
\label{dH}
\end{equation}
\end{subequations}
where the wavenumbers ${\bf k}$ and ${\bf k^\prime}$ run over the first
Brillouin zone. Notice that $\mu>d$ to ensure the convergence of~(\ref{Ek})
in the thermodynamic limit.

The key point of our qualitative arguments is as follows~\cite{Rodriguez00}. We
compare the size scaling of the typical magnitude of the scattering matrix
$(\delta{\cal H})_{\bf kk^\prime}$ with the size scaling of the level spacing
$\delta E$ in the {\it bare\/} quasiparticle spectrum of the  homogeneous
Hamiltonian ($\Delta=0$). In particular, we focus our attention on those
eigenstates laying close to the band edges ${\bf k} = 0$ (top) and ${\bf k} =
\bm{\pi}$ (bottom), where $\bm{\pi} = \pi$ and $(\pi,\pi)$ for 1D and 2D
systems, respectively. The typical fluctuation of the scattering matrix
$(\delta{\cal H})_{\bf kk^\prime}$ is $\sigma_{\mathrm{eff}} = \sigma/N^{d/2}$.
Thus, in spite of the fact that the magnitude of the disorder is $\sigma$, the
quasiparticle sees an effectively reduced value $\sigma_{\mathrm{eff}}$. It is
important that $\sigma_{\mathrm{eff}}$ scales inversely proportional to
$N^{d/2}$. Straightforward calculations of the bare energy spectrum~(\ref{Ek})
close to the band edges give the following results:
\begin{subequations}
\label{2}
\begin{equation}
E_{\bf k} \simeq E_0-J A_d(\mu)\,|{\bf k}|^{\mu-d},
\quad| {\bf k}|\to 0 \ ,
\label{0}
\end{equation}
\begin{equation}
E_{\mathbf{k}} \simeq E_{\bm{\pi}}+J B_d(\mu)\,|{\bf k}-\bm{\pi}|^2,
\quad {\bf k} \to \bm{\pi} \ ,
\label{pi}
\end{equation}
\end{subequations}
where $E_0$ and $E_{\bm{\pi}}$ are the band-edge energies and $A_d(\mu)$ and
$B_d(\mu)$ are dimensionless constants. From~(\ref{2}) it follows that the
level spacing scales as $\delta E \sim N^{-\mu+d}$  at the top of the band,
while at the bottom one gets $\delta E \sim N^{-2}$.

The matrix $(\delta {\cal H})_{\bf kk^\prime}$ couples the bare (extended)
quasiparticle states to each other and may result in their localization within
a region of size smaller than the system size. It seems reasonable to assume
that the states will be weakly coupled and consequently will be  delocalized
over the whole system provided the inequality  $\sigma_{\mathrm{eff}} \ll
\delta E$ holds. It is remarkable that for $\mu < 3d/2$, the level spacing
$\delta E$ at the top of the band diminishes upon increasing $N$ slower than
the effective magnitude of disorder $\sigma_{\mathrm{eff}}$. Therefore, if the
coupling between bare states is weak for some finite $N$
($\sigma_{\mathrm{eff}} \ll \delta E$) then it will become even weaker upon
increasing $N$. Consequently, one may expect that the state will remain
extended in the thermodynamic limit $N\to\infty$. It is also reasonable to
assume that disorder of magnitude larger than the bare bandwidth will localize
all the states. From the above arguments we conjectured the existence of an
Anderson transition in 1D and 2D systems with diagonal disorder and {\it
nonrandom\/} long-range hopping as long as $\mu < 3d/2$. Below we provide
analytical and numerical confirmations of this conjecture.

Concerning the parabolic range of the energy spectrum (close to the bottom  of
the band), we notice that the level spacing diminishes as $N^{-2}$ upon
increasing the lattice size, i.e., faster than the effective magnitude of
disorder  $\sigma_{\mathrm{eff}}$. Now, even if $\sigma_{\mathrm{eff}} \ll
\delta E$ for a small lattice size and the states are delocalized, the above
inequality will be reverted for larger $N$, resulting finally in the
localization of those states. The same conclusion holds for both band edges
within the  nearest-neighbor approximation, where the level spacing is always
$\delta E \sim N^{-2}$.

A supersymmetric method for disorder averaging~\cite{efetov,gurus}, combined
with a renormalization group (RG) analysis, provide support to the above
arguments. In short (the details will be published elsewhere), the sequence of
our steps is as follows. As a first step, we consider the one particle Green's
function with the fermionic partition function $Z_0$ and the bare action $S_0$
defined as
\begin{subequations}
\label{3}
\begin{equation}
Z_0 = \int \prod_{\bf n}  d \psi_{\bf n} \;
d \bar{\psi}_{\bf n}
\; e^{-S_0} = {\rm det} \; ( {\cal H} - {\cal E}I) \ ,
\label{Z_0}
\end{equation}
\begin{equation}
S_0 = \sum_{{\bf nm}} i \; \bar{\psi}_{\bf n}
\; ( {\cal H} - {\cal E}I )_{{\bf nm}}\;  \psi_{\bf m} \ ,
\label{S_0}
\end{equation}
\end{subequations}
where ${\cal E} = E + i 0^+$ and $I$ is the identity matrix. Introducing
bosonic ghosts $\beta, \bar{\beta}$ and expressing $1/Z_0$ as a path
integral~\cite{gurus}, we then average the  one particle Green function using
the Gaussian probability distribution of site energies, $P(\varepsilon_{\bf n})
= (1/\pi g)^{1/2} \exp(- \varepsilon_{\bf n}^2/g)$, instead of the box
distribution introduced in the beginning. This allows us to perform the
integration over site energies explicitly. The effective action
\begin{eqnarray}
S_{\mathrm {eff}} &=& i\sum_{\bf nm} \left[
\bar{\psi}_{\bf n} ( J_{\bf nm} - {\cal E} \delta_{\bf nm} )\psi_{\bf m}
\right.
\nonumber\\
&+& \left.\bar{\beta}_{\bf n} (J_{\bf nm} - {\cal E} 
\delta_{\bf nm})\beta_{\bf m} 
\right]
\nonumber\\
&+& \; \frac{g}{4} \sum_{\bf n} ( \bar{\psi}_{\bf n} \; \psi_{\bf n}
+ \bar{\beta}_{\bf n} \; \beta_{\bf n} )^2 \ ,
\label{Seff}
\end{eqnarray}
which appears after averaging, will be the main object of our RG analysis.
For doing this, it is convenient to rewrite the action in the
${\bf k}$-representation and to regroup the terms as follows
$S_{\mathrm {eff}} = S_{\mathrm {kin}}  + S_{\cal E} + S_{g}$, where
\begin{subequations}
\label{4}
\begin{eqnarray}
S_{\mathrm {kin}}  = & - & i J A_d \; \int d^d {\bf k} \; |{\bf k}|^{\mu - d}
[ \bar{\psi}({\bf k}) \; \psi({\bf k}) \nonumber\\
& + & \bar{\beta}({\bf k}) \; \beta({\bf k})] \ ,
\label{a13}
\end{eqnarray}
\begin{equation}
S_{\cal E}  =  - i {\cal E} \int d^d {\bf k} \;
[ \bar{\psi}({\bf k}) \; \psi({\bf k}) + \bar{\beta}({\bf k}) \;
\beta({\bf k})] \ ,
\label{a14}
\end{equation}
\begin{eqnarray}
S_{g} & = & \frac{g}{8 \pi} \int \prod_{i=1}^4  d^d {\bf k_i}  \;
\delta^{}({\bf k_1} + {\bf k_2} - {\bf k_3} - {\bf k_4})
\nonumber\\
&  & \times [ \bar{\psi}({\bf k_1}) \; \psi({\bf k_2})
+ \bar{\beta}({\bf k_1}) \; \beta({\bf k_2})]
\nonumber\\
&  & \times [ \bar{\psi}({\bf k_3}) \; \psi({\bf k_4})
+ \bar{\beta}({\bf k_3}) \; \beta({\bf k_4})] \ .
\label{a15}
\end{eqnarray}
\end{subequations}
We have absorbed the constant $E_0$ into ${\cal E}$ in~(\ref{a14}). Note as
well that the integration over momenta are restricted to a  $d$ dimensional
sphere of radius $\Lambda$, i.e. $|{\bf k}| < \Lambda$, with $\Lambda$ being
an ultraviolet cutoff.

The action~(\ref{4}) is the starting point of our RG analysis, which is
inspired by  Shankar's approach~\cite{shankar} to fermionic condensed matter
systems. The key observation is that the kinetic part of the  action,
$S_{\mathrm {kin}}$, is invariant under the following scaling transformation
of the cutoff $\Lambda$, the momenta ${\bf k}$ and the fields:
\begin{subequations}
\label{5}
\begin{equation}
\Lambda \rightarrow \Lambda' = \Lambda/b ,
\qquad
{\bf k} \rightarrow {\bf k}^\prime = b \; {\bf k}, \qquad b > 1
\label{a16}
\end{equation}
\begin{equation}
 \phi({\bf k}^\prime/b) = b^{\mu/2} \; \phi({\bf k}^\prime) , \qquad
\phi = \psi, \bar{\psi}, \beta, \bar{\beta} \ .
\label{a17}
\end{equation}
\end{subequations}%
For generic values of $\mu > d$, Eq.~(\ref{a16}) is a non standard 
scaling law which emerges from the unusual kinetic term~(\ref{a13}).

Driven by~(\ref{5}), the {\it mass\/} term $S_{\cal E}$ also transforms into
itself with a new {\it coupling constant\/} ${\cal E}^\prime$ given by ${\cal
E}^\prime = b^{\mu - d} \; {\cal E}$. Thus ${\cal E}$ is a relevant
perturbation of the free action $S_{\rm kin}$, as it is always the case of mass
terms~\cite{shankar}.

The term $S_g$ also transforms onto itself under the following RG 
transformation $g' = b^{2\mu-3d}\;g$. This equation implies that $g$ eventually
goes to zero upon increasing $b$ provided that $ \mu < 3d/2$. Hence, randomness
vanishes in the  low energy effective theory for $\mu < 3d/2$. On the contrary,
$g$ runs to stronger coupling whenever $ \mu > 3d/2$. 

Finally, the coupling $g$ is marginal at tree level for $\mu = 3d/2$,
and one has to consider the one loop effects to see its fate. This can 
be done using the techniques developed in Ref.~\cite{shankar}. Let
us present our main results. The RG flows of the constants $g$ and ${\cal E}$, 
up to one loop, are given by
\begin{subequations}
\label{6}
\begin{eqnarray}
\frac{d \bar{g}}{d s} & = & ( 2 \mu - 3 d)\; \bar{g}
+ \bar{g}^2 \ ,
\label{a20} \\
\frac{ d \bar{\cal E} }{d s} & = & (\mu - d) \; \bar{\cal E} - \bar{g} \ ,
\label{a21}
\end{eqnarray}
\end{subequations}%
where $s$ is the RG parameter defined as $b = \exp(s)$, and $\bar{g}$
and $\bar{\cal E}$ are related to $g$ and ${\cal E}$ as follows
\begin{equation}
g = \frac{\pi}{\Omega_d} (J A_d)^2 \; \Lambda^{2 \mu - 3 d}  \; \bar{g},
\qquad
{\cal E }= \frac{1}{4} J A_d \; \Lambda^{\mu -d} \;  \bar{\cal E}\ ,
\label{a22}
\end{equation}
with $\Omega_d$ being the volume of the $d$-dimensional sphere.

\begin{figure}[h]
\includegraphics[width=80mm]{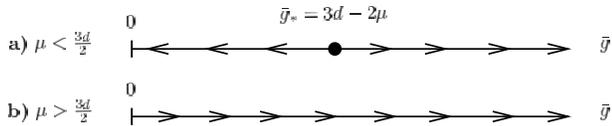}
\caption{RG-flows for the coupling constant $\bar{g}$, 
Eqs.~(\protect\ref{a20}) and~(\protect\ref{a22}), depending on the value of the
exponent $\mu$.}
\label{fig1}
\end{figure}

Equation~(\ref{a20}) has an unstable fixed point $\bar{g}_{*} = 3d - 2 \mu$
provided that $\mu < 3d/2$ (see Fig.~\ref{fig1}a). Below this point, the
coupling $g$ goes to zero asymptotically, while above $g$ grows. For $\mu >
3d/2$ the critical point disappears and the system always flows to strong
coupling (see Fig.~\ref{fig1}b). This signals about some changes in the density
of states and the different nature of the eigenfunctions when passing from $\mu
< 3d/2$ to $\mu > 3d/2$. We would like to remark the fact that the critical
value $\mu = 3d/2$ appeared in the present RG analysis coincides with that
found on the basis of our qualitative arguments.

It is usually argued that the averaged one-particle Green function does not
carry information about the spatial extent of the eigenfunctions. Therefore, we
chose the generalized inverse participation ratios (GIPR) as the relevant
quantities to characterize localization properties of the states. The standard
definition reads $I_\nu^{(q)} = \langle \sum_{\mathbf{n}}\>
|\Psi_{\nu\mathbf{n}}|^{2q} \rangle$, where $\Psi_{\nu \bf n}$ is the
probability amplitude of the normalized eigenstate $\nu$ at site $\mathbf{n}$,
and the brackets denote disorder averaging. As is well known, the GIPR scales
as $I_\nu^{(q)} \sim {\cal N}^{-q+1}$ for delocalized states ($q\neq 1$),
while shows no size scaling for localized ones. 

Our main steps in calculating the GIPR are as follows. We estimated the
uppermost eigenfunction, $\Psi_{0 \bf n}$, using perturbation theory with
respect to the random term $\sum_{\bf {kk^\prime}} (\delta {\cal H})_{\bf
{kk^\prime}}|{\bf k} \rangle \langle {\bf k^\prime}|$ in~(\ref{Hk}), namely
considering the on-site energies small compared to the hopping parameter $J$,
$|\varepsilon_{\bf n}|\ll J$. Then, $\Psi_{0\bf n}$ can be written in real
space as
\begin{subequations}
\label{7}
\begin{eqnarray}
\Psi_{0 \bf n} &=& \frac{e^{\phi_{\bf n}}}
{(\sum_{\bf n}e^{2\phi_{\bf n}})^{1/2}} \ ,\\
\phi_{\bf n} &\equiv& \sum_{\bf m}S_{\bf nm}\varepsilon_{\bf m}\ ,\\
S_{\bf nm} &\equiv &\frac{1}{{\cal N}} \sum_{\bf k\neq 0}
\frac{e^{i{\bf k}(\bf m-n)}}{E_0-E_{\bf k}} \ .
\end{eqnarray}
\end{subequations}
In order to compute the GIPR we again made use of the the supersymmetric method
for disorder averaging (as described above) with the bare action given by
Eq.~(\ref{Seff}), as well as the replica trick introduced in
Ref.~\onlinecite{ludwig}. The latter reads
\begin{equation}
I_0^{(q)} = \lim_{r\rightarrow 0} \frac{1}{N^d}
\sum_{{\bf n},{\bf n}_1,\ldots,{\bf n}_{r-q}}
\left\langle
e^{2q\phi_{\bf n}} \prod_{j=1}^{r-q} e^{2\phi_{{\bf n}_j}} \right\rangle\ .
\label{replica}
\end{equation}
In doing so, we found that $I_{0}^{(q)}\sim N^{-(q-1)d}$ provided when
$d<\mu<3d/2$; in other words, the generalized dimension equals the space
dimension so that the uppermost state is delocalized, in full agreement with
our qualitative picture.  In particular, notice that the so called inverse
participation ratio (IPR) scales as $I_0^{(2)} \sim {\cal N}^{-1}$.

\begin{figure}[ht]
\includegraphics[width=70mm,clip=]{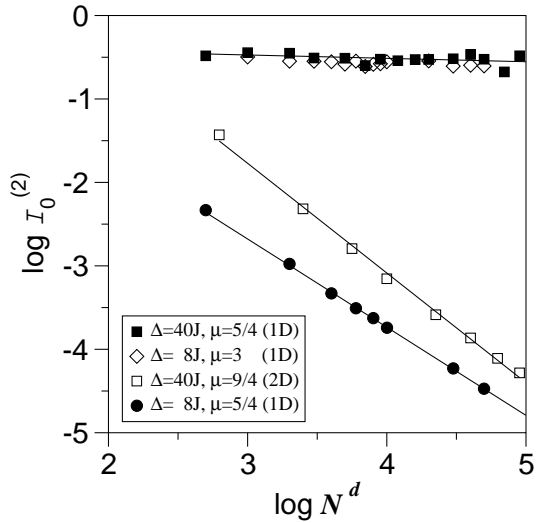}
\caption{Scaling of the IPR of the uppermost eigenstate as a function of the
number of sites ${\cal N}=N^{d}$ for different magnitudes of disorder $\Delta$
in 1D $(d = 1)$ and 2D $(d = 2)$ lattices.}
\label{fig2}
\end{figure}

Since the previous analytical study of the GIPR was perturbative, we have also
carried out a numerical study of the model to support the validity of our
conclusions. We took advantage of the Lanczos method~\cite{Golub96} as well as
the density matrix renormalization group approach~\cite{Delgado99}, allowing
one to calculate some few eigenstates of the Hamiltonian~(\ref{Hsite}) for
rather large system size. In Fig.~\ref{fig2} we plotted the IPR ($q=2$) of the
uppermost state as a function of the system size ${\cal N} = N^d$ for different
degrees of disorder $\Delta$. The behaviour of the other top states is similar
to that which manifests the uppermost state. Observing Fig.~\ref{fig2} we
conclude that the uppermost state is delocalized even for a moderately high
value of the degree of disorder ($\Delta=8J$ in 1D systems and $\Delta=40J$ in
2D system), provided $d<\mu<3d/2$. For comparison, the 1D (2D) bandwidth for
$\mu=5/4$ ($\mu=9/4$) in the absence of disorder is of the order of $10.5J$
($28J$). However, for large degree of disorder the IPR remains constant on
increasing the system size (see Fig.~\ref{fig2} for $\mu=5/4$, $d=1$ and
$\Delta=40J$), indicating that the uppermost eigenstate is localized.
Therefore, the top eigenstates undergo the Anderson transition on increasing
$\Delta$ whenever $d<\mu<3d/2$. It is to be noticed the absence of scaling of
the IPR and the subsequent localization for $d=1$, $\mu=3$, and $\Delta=8J$.
This result is in full correspondence with the  analytical analysis stating
that no transition is expected in this case. 

In summary, we have shown analytically and numerically that a power-like {\it
nonrandom\/} intersite hopping, $J_{\bf nm} = J/|{\bf n - m}|^{\mu}$, may act
towards delocalization of quasiparticle states in low dimensional systems. In
particular, the states of the top of the band may be delocalized at rather high
magnitudes of disorder ($\Delta \gg J$) and undergo a
localization-delocalization transition as the magnitude of disorder increases.
Scaling arguments provide a clear physical picture of the underlying mechanism
responsible for the delocalization of the states, in spite of the 
low-dimensional ($d\leq 2$) geometry of the system. The different size scaling
of the bare level spacing, $\delta E \sim N^{-\mu+d}$, and the magnitude of
disorder seen by a quasiparticle, $\sigma_{\mathrm{eff}} \sim \Delta/N^{d/2}$,
is the feature of the model responsible for this unusual behavior. $\delta E$
decreases slower than $\sigma_{\mathrm{eff}}$ upon increasing the system size
as long as $d < \mu < 3d/2$, resulting in the delocalization of the
corresponding quasiparticle states in the thermodynamic limit.  We stress that
the main finding of our study, namely the existence of the Anderson transition
in a physically relevant model, has been concluded on the basis of three
different approaches and the conclusions obtained are self-consistent. Most
important, the validity of the scaling analysis is not limited to the present
model. Indeed, it is established on solid grounds that the standard, three
dimensional Anderson model manifests the localization-delocalization transition
at the band center. Within this model, the bare level spacing at the band
center diminishes proportionally to $N^{-1}$, while the magnitude of effective
disorder goes down faster, $\sim N^{-3/2}$, thus being unable to localize the
states at the band center for a moderate disorder. A strong disorder (large
compared to the band width) localizes the states, giving rise a
localization-delocalization transition. We can then be confident that this kind
of scaling arguments may provide physical insight in several localization
problems.

\begin{acknowledgments}

V.~A.~M. acknowledges support from  MECyD (Project SAB2000-0103). A.~R. and
F.~D-A. were supported by DGI-MCyT (Project MAT2000-0734) and CAM (Project
07N/0075/2001). G.~S. and M.~A. M-D. acknowledge support from PGC (Project
BFM2000-1320-C02-01).

\end{acknowledgments}


\begin{thebibliography}{99}

\bibitem{Anderson58} P.\ W.\ Anderson, Phys.\ Rev.\ {\bf 109}, 1492 (1958).

\bibitem{Abrahams79} E.\ Abrahams, P.\ W.\ Anderson, D.\ C.\ Licciardello,
        and V.\ Ramakrishnan, Phys.\ Rev.\ Lett.\ {\bf 42}, 673 (1979).

\bibitem{Lee85} P.\ A.\ Lee and T.\ V.\ Ramakrishnan, Rev.\ Mod.\ Phys.\
        {\bf 57}, 287 (1985).

\bibitem{Kramer93} B.\ Kramer and A.\ MacKinnon, Rep.\ Prog.\ Phys.\
        {\bf 56}, 1469 (1993).

\bibitem{Logan87} D.\ E.\ Logan and P.\ G.\ Wolynes, Phys.\ Rev.\ B
        {\bf 29}, 6560 (1984); {\em ibid.} {\bf 31}, 2437 (1985);
        {\em ibid.} {\bf 36}, 4135 (1987); J.\ Chem.\ Phys.\
        {\bf 87}, 7199 (1987).

\bibitem{Levitov89} L.\ S.\ Levitov, Europhys. Lett. {\bf 9}, 83 (1989);
         Ann.\ Phys.\ (Leipzig) {\bf 8}, 507 (1999).

\bibitem{Mirlin96} A.\ D.\ Mirlin, Y.\ V.\ Fyodorov, F.-M.\ Dittes, J.\ Quezada,
        and T.\ H.\ Seligman, Phys. Rev. E {\bf 54}, 3221 (1996).

\bibitem{Parshin98} D.\ A.\ Parshin and H.\ R.\ Schober, Phys.\ Rev.\ B 
        {\bf 57}, 10232 (1998).

\bibitem{Deych00} L.\ I.\ Deych, A.\ A.\ Lisyansky, and B.\ L.\ Altshuler,
	Phys.\ Rev.\ Lett. {\bf 84}, 2678 (2000);
	Phys.\ Rev.\ B {\bf 64}, 224202 (2001).

\bibitem{Nabetani95} A.\ Nabetani, A.\ Tamioka, H.\ Tamaru, and K..\ Miyano,
        J. Chem. Phys. {\bf 102}, 5109 (1995).

\bibitem{Kopelman97} R.\ Kopelman, M.\ Shortreed, Z.-Y.\ Shi, W.\ Tan, Z.\ Xu,
        J.\ Moore, A.\ Bar-Haim, and J.\ Klafter, Phys. Rev. Lett. {\bf 78},
        1239 (1997).

\bibitem{dendrimers}  M.\ A.\ Mart\'{\i}n-Delgado, 
        J.\ Rodr\'{\i}guez-Laguna, G.\ Sierra,
        Phys.\ Rev.\ {\bf B 65}, 155116 (2002); cond-mat/0012382.

\bibitem{Rodriguez00} A.\ Rodr\'{\i}guez, V.\ A.\ Malyshev, and F.\
        Dom\'{\i}nguez-Adame, J.\ Phys.\ A: Math.\ Gen.\ {\bf 33}, L161
	(2000).

\bibitem{efetov} K.\ B.\ Efetov, Adv.\ Phys.\ {\bf 32}, (1983) 53.

\bibitem{gurus}  S.\ Guruswamy, A.\ LeClair, A.\ W.\ W.\ Ludwig,
        Nucl.\ Phys.\ {\bf B583} (2000) 475.

\bibitem{shankar} R.\ Shankar, Rev.\ Mod.\ Phys.\ {\bf 66} (1994) 129.

\bibitem{ludwig} A.\ W.\ W.\ Ludwig, M.\ P.\ A.\ Fisher, R.\ Shankar, and G.\
	Grinstein, Phys.\ Rev.\ {\bf B 50}, 7526 (1994).

\bibitem{Golub96} G.\ H.\ Golub and C.\ F.\ Van Loan, {\em Matrix
        Computations\/} (The Johns Hopkins University Press, Maryland, 1996).

\bibitem{Delgado99} M.\ A. Mart\'{\i}n-Delgado, G.\ Sierra, and R.\ M.\ Noack,
        J.\ Phys.\ A: Math.\ Gen.\ {\bf 32} 6079 (1999).

\end{thebibliography}
\end{document}